\documentclass[12pt]{article}
\usepackage{epsfig}
\usepackage{graphicx}
\usepackage[space,nosort]{cite}

\def\Title#1{\begin{center} {\Large {\bf #1} } \end{center}}
\usepackage{fancyhdr}
\begin{document}

\begin{flushright}\small
CERN-PH-TH/2012-105,\ \
SLAC-PUB-14956,\ \
IPPP/12/17,\ \ DCPT/12/34,\ \ LPN12-045
\end{flushright}
\vskip7mm
\Title{Systematic improvement of QCD parton showers}
\bigskip

\begin{raggedright}  

{\it\underline{Jan Winter} \index{}
\footnote{Presented at Linear Collider 2011: Understanding QCD at Linear Colliders  in searching for old and new physics, 12--16 September 2011, ECT*, Trento, Italy}\\
PH-TH Department, CERN, CH-1211 Geneva 23, Switzerland\\
{\rm jwinter@cern.ch}
}\\
\vspace{0.2cm}
{\it Stefan H\"oche \index{}\\
Theory Department, SLAC National Accelerator Laboratory, Menlo Park,
CA 94025, USA\\
{\rm shoeche@slac.stanford.edu}
}\\
\vspace{0.2cm}
{\it Hendrik Hoeth, Frank Krauss, Marek Sch\"onherr and Korinna Zapp \index{}\\
IPPP, Durham University, Durham DH1 3LE, UK\\
{\rm hendrik.hoeth@durham.ac.uk, frank.krauss@durham.ac.uk, marek.schoenherr@durham.ac.uk, k.c.zapp@durham.ac.uk}
}\\
\vspace{0.2cm}
{\it Steffen Schumann \index{}\\
II. Physikalisches Institut, Universit\"at G\"ottingen,
D-37077 G\"ottingen, Germany\\
{\rm steffen.schumann@phys.uni-goettingen.de}
}\\
\vspace{0.2cm}
{\it Frank Siegert \index{}\\
Physikalisches Institut, Albert-Ludwigs-Universit\"at Freiburg,
D-79104 Freiburg, Germany\\
{\rm frank.siegert@cern.ch}
}

\bigskip\bigskip
\end{raggedright}
\vskip 0.5  cm
\begin{raggedright}
{\bf Abstract} In this contribution, we will give a brief overview of
the progress that has been achieved in the field of combining matrix
elements and parton showers. We exemplify this by focusing on the case
of electron--positron collisions and by reporting on recent
developments as accomplished within the
S\protect\scalebox{0.77}{HERPA} event generation framework.
\end{raggedright}

\newcommand{\Sherpa}{S\protect\scalebox{0.77}{HERPA}}
\newcommand{\Pythia}{P\protect\scalebox{0.77}{YTHIA}}
\newcommand{\Herwig}{H\protect\scalebox{0.77}{ERWIG}}
\newcommand{\CSshower}{CS\protect\scalebox{0.77}{SHOWER}}
\newcommand{\METS}{ME\protect\scalebox{0.77}{\&}TS}
\newcommand{\MCNLO}{MC\protect\scalebox{0.77}{@NLO}}
\newcommand{\POWHEG}{P\protect\scalebox{0.77}{OWHEG}}
\newcommand{\MENLOPS}{ME\protect\scalebox{0.77}{NLO}PS}

\section{Monte Carlo event generation at a glance\label{sec:mc}}
Event generators are widely used to model the multi-hadron final
states of high-energy particle collisions. For a very comprehensive
review, we refer the interested reader to Ref.~\cite{Buckley:2011ms}.
The underlying principle for organizing the computer simulation of
events is factorization, i.e.\ to factorize the evolution of each
event into several phases ordered according to their energy domains.
We broadly distinguish two major phases governed by two different
physics regimes: we can apply short-distance/perturbative methods to
describe the physics at the harder energy scales while for the
description of soft effects, we have to rely on phenomenological
models encoding our observations regarding the confinement of the
collision products, a mechanism for which a rigorous understanding has
not been developed yet. The separation is mainly driven by the nature
of QCD where the strong coupling becomes small at large scales, such
that the theory becomes asymptotically free and can be formulated in
terms of partons. Contrary at scales of ${\cal O}(1)$ GeV, the
coupling strength has increased substantially and non-perturbative
dynamics dictates the evolution of the events. An extremely important
property of QCD is the formation of jets, which manifest themselves as
sprays of particles leaving localized energy deposits in the detectors.
Correspondingly, the phases of the event generation can also be
described in terms of jet production and (intra-jet) evolution, cf.\
e.g.\ Ref.~\cite{Winter:2007zz}.

\begin{figure}[t!]
  \vskip-3mm
  \begin{center}
    \resizebox{0.37\textwidth}{!}{
      \includegraphics{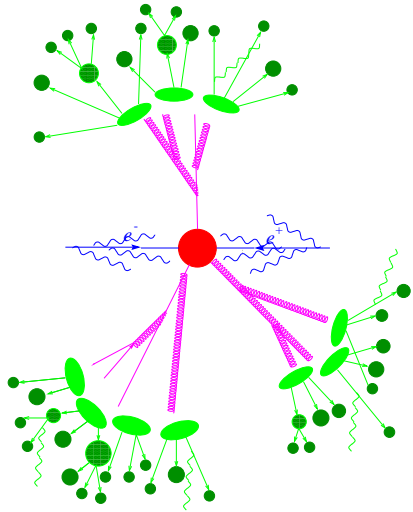}}
    \vskip-7mm
    \caption{The various phases of Monte Carlo event generation,
      illustrated for lepton--lepton collisions. The outer, circular
      part visualizes the event evolution driven by non-perturbative
      dynamics (depicted by the green blobs) while the inner part
      shows the phases related to short-distance phenomena (depicted
      by the red and blue objects).}
    \label{fig:stages}
  \end{center}
  \vskip-5mm
\end{figure}

Fig.~\ref{fig:stages} gives the details by showing not only the main
but all phases, which we consider in Monte Carlo event generation. The
phases where physics can be mastered with perturbative methods are
visualized in the inner part of the figure. In blue we show the
effects of initial-state radiation off the incoming leptons, which
commonly are encoded in an inclusive manner by electron structure
functions. The red objects visualize the hard interaction (shown by
the big red blob in the middle representing the process $e^-e^+\to q\bar qg$)
producing energetic parton jets that give rise to subsequent QCD
bremsstrahlung (shown by the branching pattern in magenta). The
physics of the hard process is best described by relying on exact
matrix-element expressions -- with the current frontier given by
$n$-leg tree-level ($n\sim10$) and QCD virtual ($n\sim5$) amplitudes
-- whereas all bremsstrahlung effects are described by parton
showering based on matrix-element approximations that are correct in
the singular phase-space regions of QCD. The phases of
non-perturbative dynamics are represented by the green-coloured blobs
in the outer sphere of the figure. They depict the transition of the
coloured partons into primary, unstable hadrons and their subsequent
decays into stable, detectable hadrons, which can be described by
phase-space or effective models. The parton--hadron conversion is
``parametrized'' by hadronization models, such as the renowned models
of Lund string fragmentation \cite{Andersson:1983ia} or cluster
hadronization \cite{Webber:1983if}.

A full-fledged Monte Carlo event generator incorporates physics
implementations according to all phases of event evolution, from the
evaluation of scattering matrix elements to the description of hadron
decays. The Monte Carlo approach is inherent to all phases: cross
sections are physical objects and, hence, a probabilistic picture can
be identified for each phase. We can draw events from the resulting
probability densities by generating random
numbers. \Pythia~\cite{Sjostrand:2006za},
\Herwig~\cite{Corcella:2000bw,Corcella:2002jc} and
\Sherpa~\cite{Gleisberg:2003xi,Gleisberg:2008ta} are examples for
(well) established event generators in the LHC era. Common to them is
the generation of hadron-level predictions, which can be compared
directly to experimental data, once the data are corrected for
detector effects.

\section{Parton shower basics and modern formalisms\label{sec:ps}}

The final states of the hard interactions often produce partons that
are still sufficiently energetic to induce further radiation, because
there is enough time for them to interact perturbatively before
hadronization sets in. Owing to the singularity structure intrinsic to
QCD, these emissions preferably populate the collinear and soft
regions of phase space, and very conveniently it is in these limits
that QCD amplitudes factorize. This can be taken further, i.e.\
be promoted to a factorization at the cross-section level:
\begin{equation}\label{eq:xsec.recursion}
d\sigma_{n+1}\;=\;d\sigma_n\;
\frac{\alpha_s(t)}{2\pi}\;\frac{dt}{t}\;dz\;P_{a\to bc}(z)\ .
\end{equation}
Here $\alpha_s$, $t\equiv p^2_a$ and $z$\/ respectively denote the
strong coupling constant, the propagator and the momentum-fraction
variable used in the splitting process. The function $P_{a\to bc}(z)$
characterizes the parton splitting $a\to bc$ (e.g.\ $q\to qg$) in
detail, encoding the functional dependence on $z$, and possibly the
splitting angle. For example, if one considers the leading collinear
region, i.e.\ small-angle radiation off outgoing partons, the
Altarelli--Parisi (or DGLAP) splitting functions are obtained; a nice
introduction to the subject can be found in \cite{Ellis:1991qj}.
Eq.~(\ref{eq:xsec.recursion}) expresses more than factorization of the
multi-parton cross section, it ultimately forms the basis for a
recursive definition of multiple emissions ordered in $t$. As a result
collinear/soft partons can be added in an iterative procedure, and we
arrive at an emission pattern as shown in Fig.~\ref{fig:stages} where
the initially energetic $q\bar qg$\/ partonic ensemble has evolved
down to a scale (magnitude of the ordering variable $t$) of the order
of $t^{1/2}\sim1$~GeV. This (i) regulates the (collinear) divergences
and (ii) sets a scale conveniently close to the onset of
hadronization. Emissions below this cut-off are said to be
unresolvable. The iterative scheme ensures that all kinematically
enhanced contributions are taken into account, which from a more
formal point of view means that the leading logarithmic (LL) terms are
summed up to all orders. The enhancements are manifest in the
intra-jet evolution and in the rapid particle multiplicity growth,
both of which being well described by the parton shower approximation.

Over the last decade the activities in the field of parton shower
modelling have been seen to be intensified for several reasons; there
was a push for designing new Monte Carlo programs for the LHC era
resulting in a careful revision of existing programs.\footnote{The
  next-generation programs \Pythia8~\cite{Sjostrand:2007gs} and
  \Herwig++~\cite{Bahr:2008pv,Gieseke:2011na} emerged from this
  initiative.} There was also a strong demand to
adjust parton showers to work well with input from (multi-leg and
loop) higher-order matrix elements, and furthermore to interconnect
them with models for multiple parton interactions and the underlying
event. These efforts led to a number of refinements in shower
algorithms and, moreover, the construction of new parton showers
\cite{Gieseke:2003rz,Sjostrand:2004ef,Nagy:2005aa,Bauer:2007ad,Nagy:2007ty,Giele:2007di,Dinsdale:2007mf,Schumann:2007mg,Winter:2007ye,Bauer:2008qh,Corke:2010zj,Giele:2011cb,GehrmannDeRidder:2011dm,Kilian:2011ka,Platzer:2012qg,Nagy:2012bt}.
We want to illustrate this very briefly by presenting two selected
results obtained from dipole-like shower schemes developed within the
\Sherpa\ collaboration.

\begin{figure}[t!]
  \begin{center}
    \resizebox{0.87\textwidth}{!}{
      \includegraphics{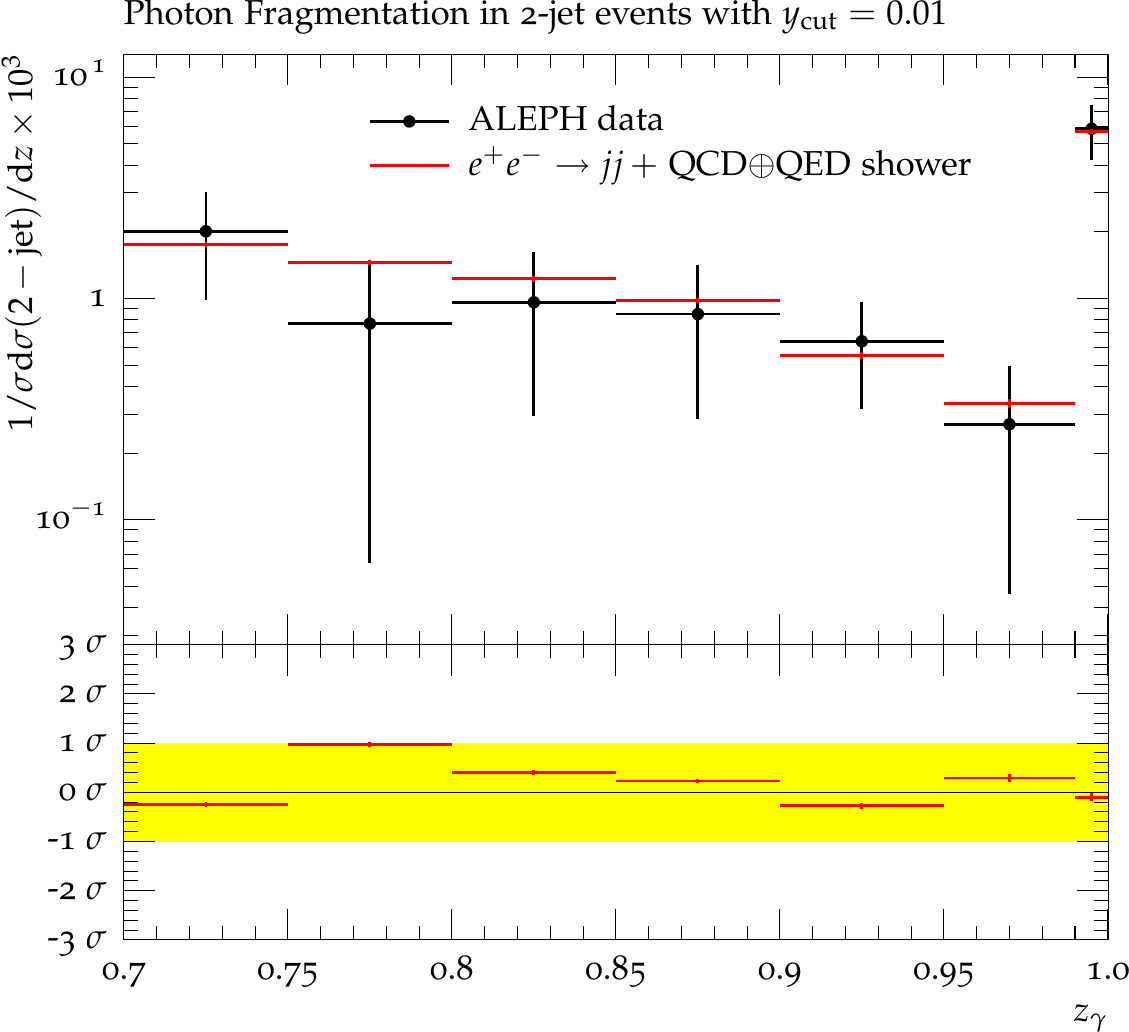}\hspace{20mm}
      \includegraphics{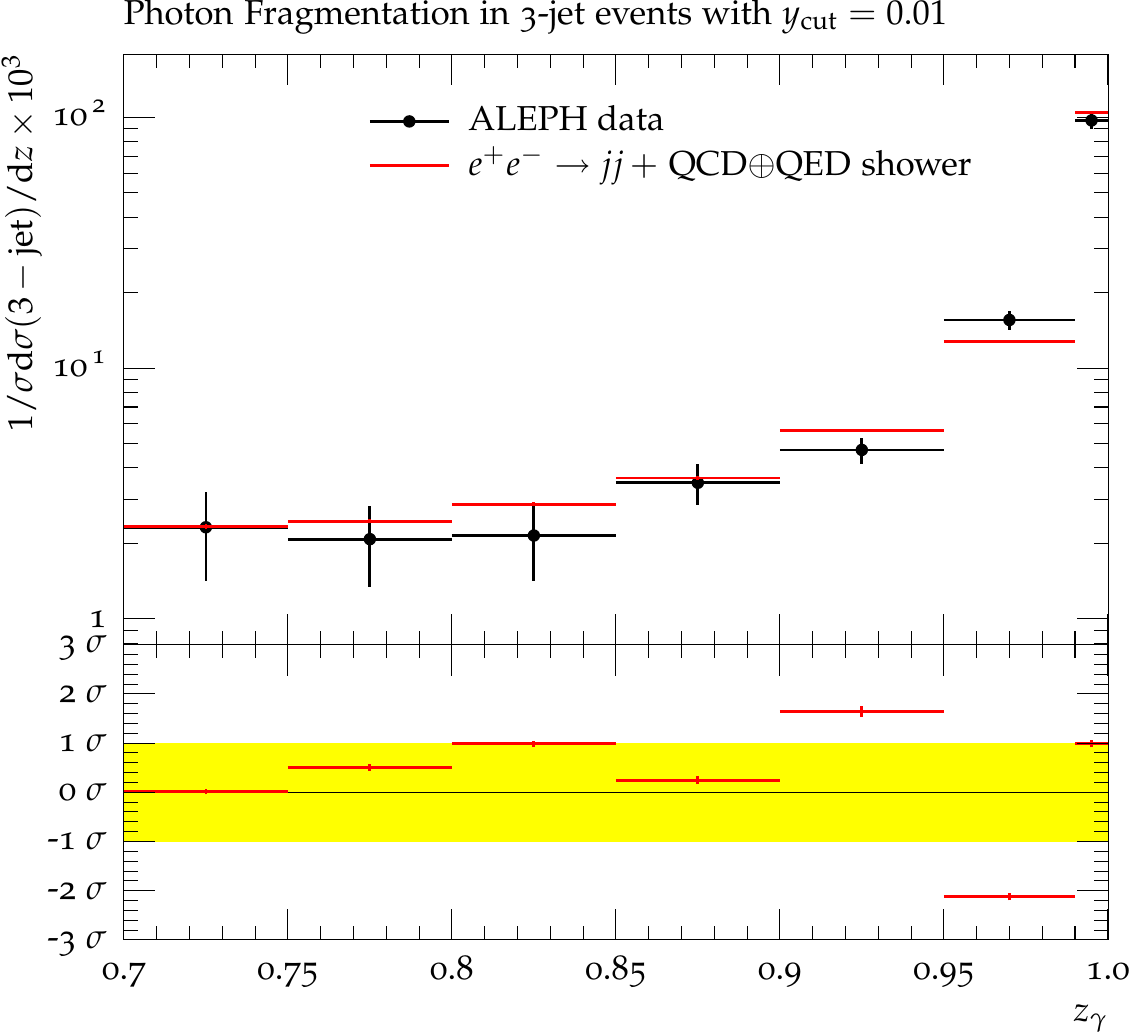}}
    \caption{The $z_\gamma$ distribution as measured by ALEPH in
      hadronic $Z^0$ decays at LEP1 \cite{Buskulic:1995au} and
      predicted by \Sherpa's QCD+QED \CSshower\ evolution added to
      $e^-e^+\to q\bar q$\/ hard scatterings.}
    \label{fig:lep.qcd+qed}
  \end{center}
  \vskip-5mm
\end{figure}

\subsection{Example -- \Sherpa's \CSshower}
The \CSshower\ was derived from the dipole subtraction formalism used
in next-to-leading order (NLO) calculations where CS stands for the
names of the pioneers of this formalism, Catani and Seymour. To
construct the shower algorithm, in particular its corresponding
splitting functions, one exploits the dipole factorization of the
real-emission matrix elements; the various CS dipole functions are
translated into shower kernels by working in 4 dimensions, the large
$N_{\mathrm C}$ limit, and averaging over spins. This was originally
described in \cite{Nagy:2005aa} and worked out in detail, as well as
implemented, in Refs.~\cite{Dinsdale:2007mf,Schumann:2007mg};
furthermore, dipole showers were verified to reproduce the DGLAP
equation \cite{Dokshitzer:2008ia,Nagy:2009re,Skands:2009tb}. The
\CSshower\ entails nice properties such as its Lorentz-invariant
formulation, on-shell splitting kinematics with rather local recoil
compensation by spectator partons, exact/complete phase-space mapping
of emissions and an inherent inclusion of soft colour coherence.
Nevertheless, for the production of vector bosons in hadronic
collisions, one (rather minor) shortcoming of the initially proposed
NLO-like recoil strategy particularly was discussed in the literature
\cite{Nagy:2009vg}. Unlike in $b$-space exponentiation this recoil
scheme does not generate the vector boson $p_T$ spectra continuously
through each emission, but finally resolutions were put forward as in
\cite{Platzer:2009jq,Hoeche:2009xc,Carli:2010cg}.

The \CSshower\ allows for the straightforward inclusion of QED effects;
technically there is almost no difference between a $q\to q\gamma$\/
and $q\to qg$\/ splitting apart from the spectator concept (all
oppositely charged particles in QED versus the colour-linked parton in
large $N_{\mathrm C}$ QCD). The respective emission probabilities
factorize trivially allowing a democratic treatment of photon and QCD
parton radiation. This has been discussed in \cite{Hoeche:2009xc}. As
an example we show in Fig.~\ref{fig:lep.qcd+qed} results of a crucial
benchmark for the combined QCD+QED \CSshower\ model, which is to
reproduce the scale-dependent photon fragmentation function
$D_\gamma(z_\gamma,y_\mathrm{cut})$ as measured by the ALEPH
collaboration in hadronic $Z^0$ decays at LEP1 \cite{Buskulic:1995au}.
The events are classified by $n$-jet topologies and resolution
measures $y_\mathrm{cut}$, and are required to have at least one
reconstructed jet containing a photon with energy fraction
$z_\gamma>0.7$ and $E_\gamma>5$~GeV. We observe a very nice agreement
between simulation and data.

\begin{figure}[t!]
  \begin{center}
    \resizebox{0.87\textwidth}{!}{\includegraphics{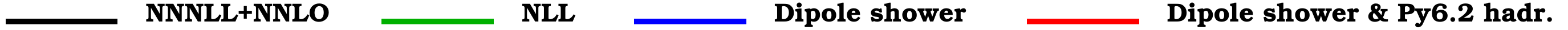}}
    \resizebox{0.87\textwidth}{!}{\includegraphics{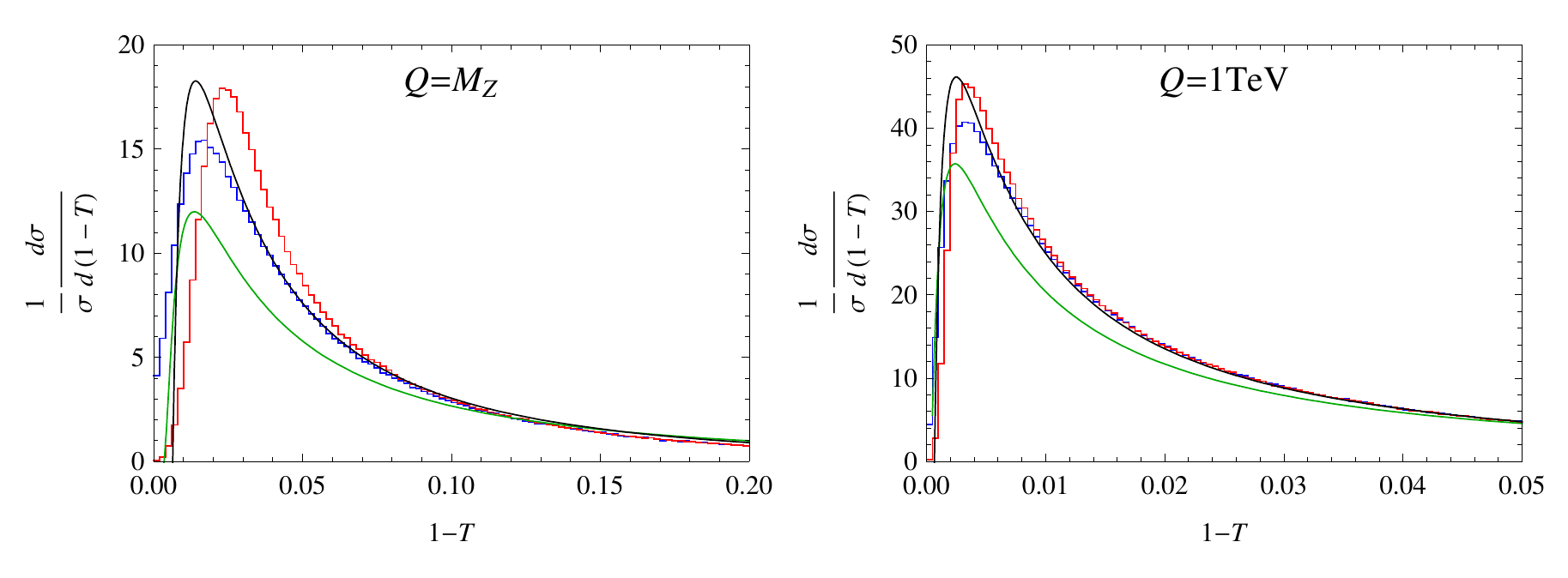}}
    \caption{The all-particle one-minus-thrust event shapes in
      electron--positron annihilation at LEP1 and TeV energies. The
      comparison is between analytical results at N$^3$LL+NNLO and NLL
      accuracy (black and green curves), and numerical results
      obtained from \Sherpa's colour dipole model neglecting/including
      hadronization corrections (blue/red histograms).}
    \label{fig:lep.dipole}
  \end{center}
  \vskip-5mm
\end{figure}

\subsection{Example -- \Sherpa's dipole shower}
While the \CSshower, incorporating $1\to2$ splittings, is said to be
dipole-like owing to the spectator involvement in constructing the
splitting kinematics, the currently unreleased shower model presented
in \cite{Winter:2007ye} is based on exploiting the QCD property of
antenna factorization in soft gluon emissions. This enables a complete
$2\to3$ treatment of the splitting process employing $2\to3$ splitting
functions and $2\to3$ kinematics. The original idea goes back to the
pioneering work of Gustafson \cite{Gustafson:1986db,Gustafson:1987rq}
resulting in the release of the successful \textsc{Ariadne} program
\cite{Lonnblad:1992tz}.

Although, as described in Ref.~\cite{Winter:2007ye}, the goal of
unifying initial- and final-state radiation into a single perturbative
framework was greatly achieved, here we only want to recall a nice
result obtained during verification of the (\textsc{Ariadne}-like)
final-state showering of \Sherpa's colour dipole model. In
Fig.~\ref{fig:lep.dipole} we display various predictions for the
all-particle $1-T$\/ distribution in $e^+e^-$ annihilation, where
$T$\/ denotes the event-shape variable thrust. By comparing directly
to theoretical results from analytic resummations at next-to-LL (NLL)
level and beyond, we obtain a stringent and unambiguous test of the
resummation as encoded in the dipole shower without the need for
hadronization corrections.\footnote{Hadronization corrections are on
  the order of $1/Q$\/ (see Fig.~\ref{fig:lep.dipole}), broaden all
  jets and shift the results towards smaller $T$\/ as seen by
  comparing the red and blue histograms.} The pure shower results turn
out to be significantly different from the NLL predictions (green
curves), they actually are closer to the N$^3$LL resummed results
\cite{Becher:2008cf}, which were calculated using soft-collinear
effective theory and matched to NNLO predictions
\cite{GehrmannDeRidder:2007bj,GehrmannDeRidder:2007hr}. This is a
rather remarkable result for the dipole shower.

\section{Higher precision for parton shower predictions\label{sec:beyond}}

Traditionally it was PSs (parton showers) that were used to describe
any additional jet activity, including the production of further hard
jets. The shower ``seeds'' are given by QCD LO processes for a fixed
final-state multiplicity. For these reasons, parton shower algorithms
are said to describe multi-jet production at the LO+LL level. But
there are a number of limitations to this description. Shower
algorithms only represent the semi-classical picture of the entire
branching process: quantum interferences and multi-parton correlations
are hardly taken into account, and the whole evolution is only
formulated in the limit of a large number of colours, $N_{\mathrm C}$.
The application of the shower approximations outside the singular
regions of QCD leads to uncontrolled behaviour and highly inaccurate
predictions for rather energetic and/or large-angle radiation; shower
uncertainties can therefore get large, and in general they are not
easy to assess, which has the potential danger of partly compensating
for missing perturbative effects via the tuning of non-perturbative
parameters.

It was clear, to systematically correct for these deficiencies, the
shower generators had to be improved by using more precise MEs (matrix
elements). Motivated by the ground-breaking advances in efficiently
calculating multi-leg MEs at tree and, more recently, even loop level,
the theoretical effort in enhancing the accuracy of PSs has resulted
in two new developments with significant impact on doing collider
phenomenology (cf.\ e.g.~\cite{Lykken:2011uv}): tree-level matrix
elements merged with parton showers (ME+PS), and NLO calculations
interfaced (or matched) with parton showers (NLO+PS). The former
primarily originated from the Catani--Krauss--Kuhn--Webber (CKKW)
paper \cite{Catani:2001cc}, with the innovative idea to correct the
first few hardest shower emissions by using exact tree-level
matrix-element expressions. A vast body of literature has appeared
subsequently, advocating several variants, implementations and
refinements to the original method (see
Refs.~\cite{Alwall:2007fs,Lavesson:2007uu,Buckley:2011ms,Maestre:2012vp}
for a review). Well-known variants include
CKKW~\cite{Catani:2001cc,Krauss:2002up,Mrenna:2003if},
L\"onnblad-CKKW~\cite{Lonnblad:2001iq,Lonnblad:2011xx},
Mangano's MLM method~\cite{Mangano:2001xp,Mangano:2006rw} and the
versions of matrix-element and truncated-shower merging
(\METS)~\cite{Hoeche:2009rj,Hamilton:2009ne}, all producing so-called
{\em improved}\/ LO+LL descriptions of multi-jet observables.

The NLO+PS development was initiated by the \MCNLO\ papers
\cite{Frixione:2002ik,Frixione:2003ei} and followed later by the
\POWHEG\ proposal \cite{Nason:2004rx,Frixione:2007vw}. Both approaches
aim at improving the event generation of a basic process in such a way
that NLO accuracy is reached for inclusive observables, while
maintaining the LL accuracy of the shower approach. Essentially, this
is achieved by raising the order of precision of the underlying core
process. In the context of multi-jet production, we then arrive at a
description accurate at NLO+LL level (see
Ref.~\cite{Nason:2012pr,Maestre:2012vp} for a very recent review). In
both cases, ME+PS and NLO+PS, we have to solve two major problems
simply because MEs and PSs can describe the same final state: the
emission phase space has to be covered in a way that double counting
of contributions is removed and dead regions are avoided at
the same time.

The theoretical effort behind these two developments has led to
enormous progress in the last decade regarding the systematic
embedding of higher-order QCD corrections in multi-purpose Monte Carlo
event generators. \Pythia, \Herwig\ and \Sherpa\ provide solutions
(partly relying on interfaces to specialized tools) and implementations
to make these developments available in experimental analyses and
collider studies. Using the new tools, we have found better agreement
to a broad range of QCD jet data taken at lepton and hadron colliders.
We have gained better control over the systematic uncertainties of the
generator predictions, and generally have been able to reduce these
uncertainties. In the remainder of this contribution, we will quickly
summarize the status of the ME+PS and NLO+PS techniques in \Sherpa.

\subsection{\METS\ in \Sherpa}
The \METS\ implementation in \Sherpa\ is state-of-the-art. Predictions
are obtained from merging tree-level matrix elements for
$X$\/ plus $0,\ldots,n$-parton final states with the \CSshower, while
preserving the LL accuracy to which soft and collinear multiple
emissions are described by the \CSshower. The new \METS\ merging
scheme was introduced in Ref.~\cite{Hoeche:2009rj} and optimized as
documented in Refs.~\cite{Hoeche:2009xc,Carli:2010cg} to improve over
the original \Sherpa\ implementation based on the CKKW
approach~\cite{Catani:2001cc,Krauss:2002up,Krauss:2004bs,Gleisberg:2005qq}.
\METS\ guarantees great compatibility between the ($Q$) scales used to
resolve the matrix-element final states and those ($t$) scales ordering
and driving the parton showering. In particular, truncated showering
has been enabled to insert important soft emissions between resolved
parton jet seeds. These shower emissions themselves do not give rise
to jets but are necessary to retain the accuracy of the shower
evolution, for example restore soft colour coherence. The very basic
steps of the \METS\ algorithm are:

\begin{figure}[t!]
  \begin{center}
    \resizebox{0.87\textwidth}{!}{
      \includegraphics{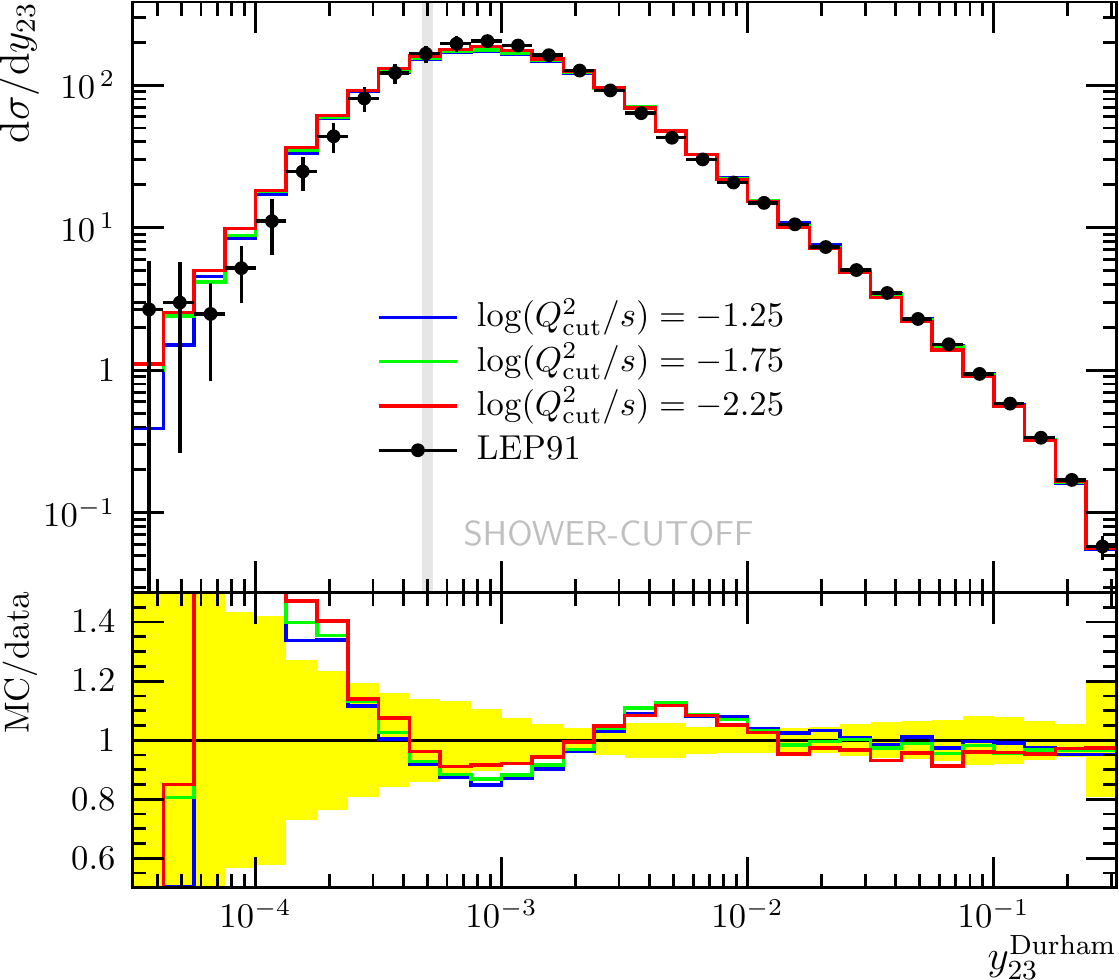}\hspace{20mm}
      \includegraphics{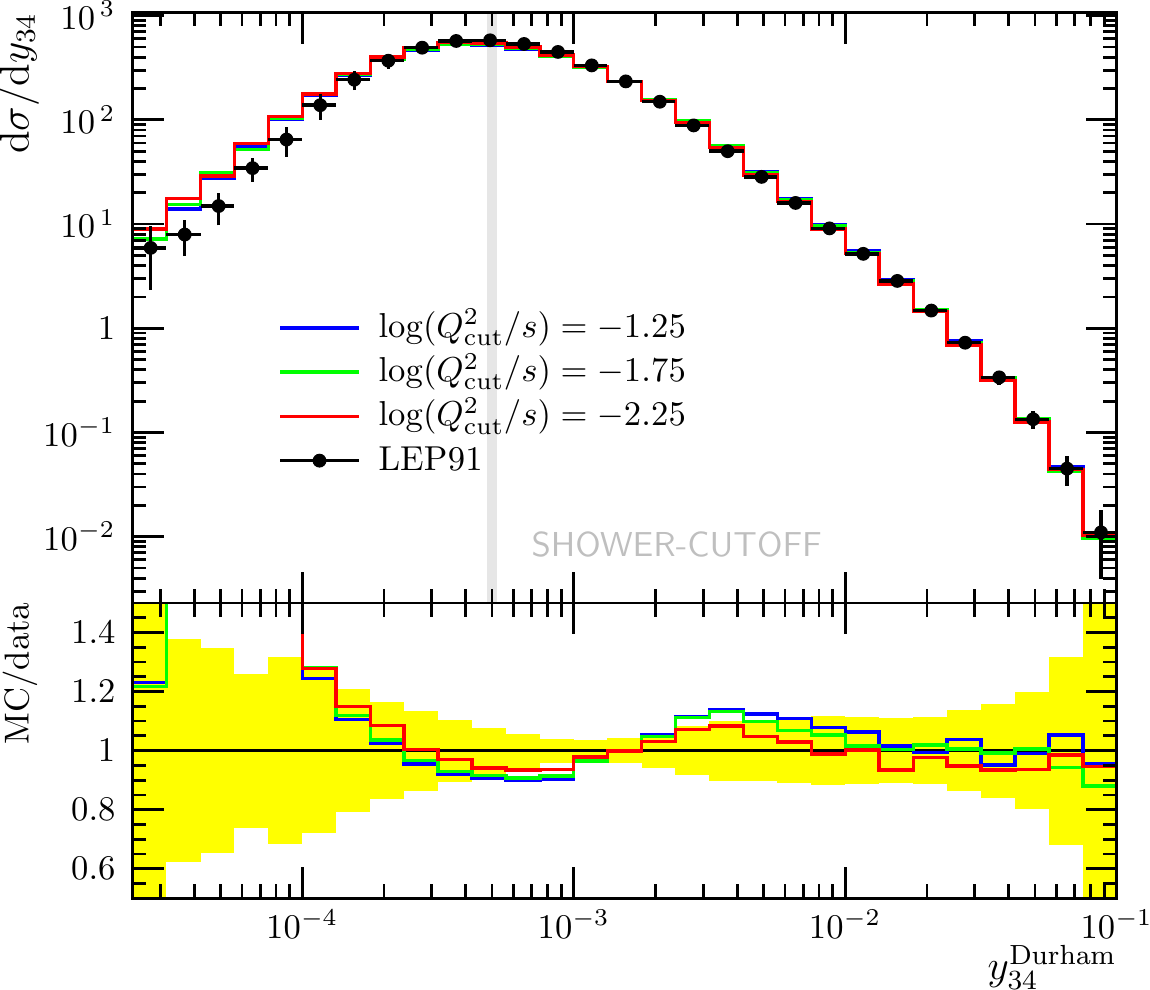}}
    \caption{Differential Durham jet rates obtained from a \Sherpa\
      \METS\ sample where up to 6 jets are described by MEs. Results
      are shown for 3 different choices of the merging-cut parameter
      and compared to LEP1 data~\cite{Pfeifenschneider:1999rz}.}
    \label{fig:lep.mets}
  \end{center}
  \vskip-5mm
\end{figure}

Separate phase space into a ``hard'' ME ($Q>Q_\mathrm{cut}$) and
``soft'' PS ($Q<Q_\mathrm{cut}$) domain according to a suitably chosen
infrared-safe jet criterion. This factorizes the shower kernels
similarly and regularizes the matrix elements. Via ``inverted''
showering one then finds the likely PS histories for the generated
$n$-parton MEs. Based on the selected history one further evolves
(using the $t$\/ scales) the ME final state beyond $n$\/ partons
unless one encounters a shower emission above $Q_\mathrm{cut}$
resulting in the rejection of the event. This way one replaces the
shower kernels in the ME domain by exact ME expressions for the
hardest $n$\/ jets, and ensures that rejected events are to be
described by ($n+1$)-parton MEs.

We exemplify the performance of \Sherpa's \METS\ merging by showing
differential jet rates for $e^+e^-$ annihilation into hadrons. The
\METS\ sample was generated by including matrix elements with up to
four extra partons ($u,d,s,c,b$ quarks and gluons). The $y_{n\,n+1}$
distributions show at which rate, according to the Durham $k_T$
algorithm, $n+1$ jets are clustered into $n$\/ jets as a function of
(the resolution parameter) $y_{n\,n+1}\approx Q^2_{n\,n+1}/s$. The
observable is very sensitive to the jet emission pattern, therefore,
lends itself eminently to assess the
$Q_\mathrm{cut}=\sqrt{s\,y_\mathrm{cut}}$ parameter dependence of the
\METS\ merging. Fig.~\ref{fig:lep.mets} shows predictions for various
$Q_\mathrm{cut}$, found to be in good agreement with data from LEP1
($\sqrt s=91.25$~GeV) \cite{Pfeifenschneider:1999rz}. Owing to the ME
inclusion the high scales are well described, while it is very
reassuring to see that the good shower behaviour is maintained at
medium scales. The low scales below the (marked) shower cut-off are
affected by hadronization effects and related parameter tuning (not
optimized here). We therefore conclude that the merging systematics is
well below the 10\% level, which is a remarkable improvement over
earlier merging variants.

\subsection{\POWHEG\ and \MENLOPS\ in \Sherpa}
The first results of a NLO+PS effort in \Sherpa\ were published in
Ref.~\cite{Hoche:2010pf}. The implementation has been based on the
\POWHEG\ formalism, which can be understood as an advancement of earlier
methods developed to correct the leading shower emission by the
corresponding real-emission ME \cite{Seymour:1994we,Miu:1998ju}. This
was done by invoking the Sudakov veto algorithm with an additional
weight to be respected, schematically written as
$w(\Phi_R)=R(\Phi_R)/R^{(\mathrm{PS})}(\Phi_R)$ where $\Phi_R$ denotes
the full real-emission phase space and $R$\/ ($R^{(\mathrm{PS})}$)
stands for the real-emission ME (shower) expression. The \POWHEG\ method
reweights similarly, but at the same time accounts for a local
$K$-factor implemented through a NLO event weight
$\bar B=B+V+I+\int d\Phi_{R|B}(R-S)$ where $\Phi_{R|B}$ is the
one-particle emission phase space. This way one generates not only
observable shapes showing the Sudakov suppression and ME improvement
at low and high scales, respectively, but also NLO accuracy for the
event sample, hence featuring a reduced scale dependence. The matching
is smooth in a sense that no phase-space cut is needed as in
(conventional) ME+PS methods.

\Sherpa\ possesses almost all ingredients that make a \POWHEG\ automation
possible: automated tree-level ME generators provide the Born ($B$)
and real-emission ($R$) terms \cite{Krauss:2001iv,Gleisberg:2008fv},
the integrated and explicit subtraction terms ($I$\/ and $S$) are
given by the automated implementation of the CS dipole subtraction
formalism \cite{Gleisberg:2007md} and the virtual contributions ($V$)
are obtained via interfacing to one-loop ME libraries as facilitated
e.g.\ by \textsc{BlackHat}~\cite{Berger:2008sj},
\textsc{GoSam}~\cite{Cullen:2011ac} or
\textsc{MCfm}~\cite{Campbell:2011bn} using the Binoth Les Houches
Accord~\cite{Binoth:2010xt}. Last but not least the \CSshower\ is well
suited for combination with the ME computations; its
$R^{(\mathrm{PS})}(\Phi_R)$ often closely approximate the $R(\Phi_R)$
resulting in a very reasonable distribution of the $w(\Phi_R)$
weights.

\begin{figure}[t!]
  \begin{center}
    \resizebox{0.87\textwidth}{!}{
      \includegraphics{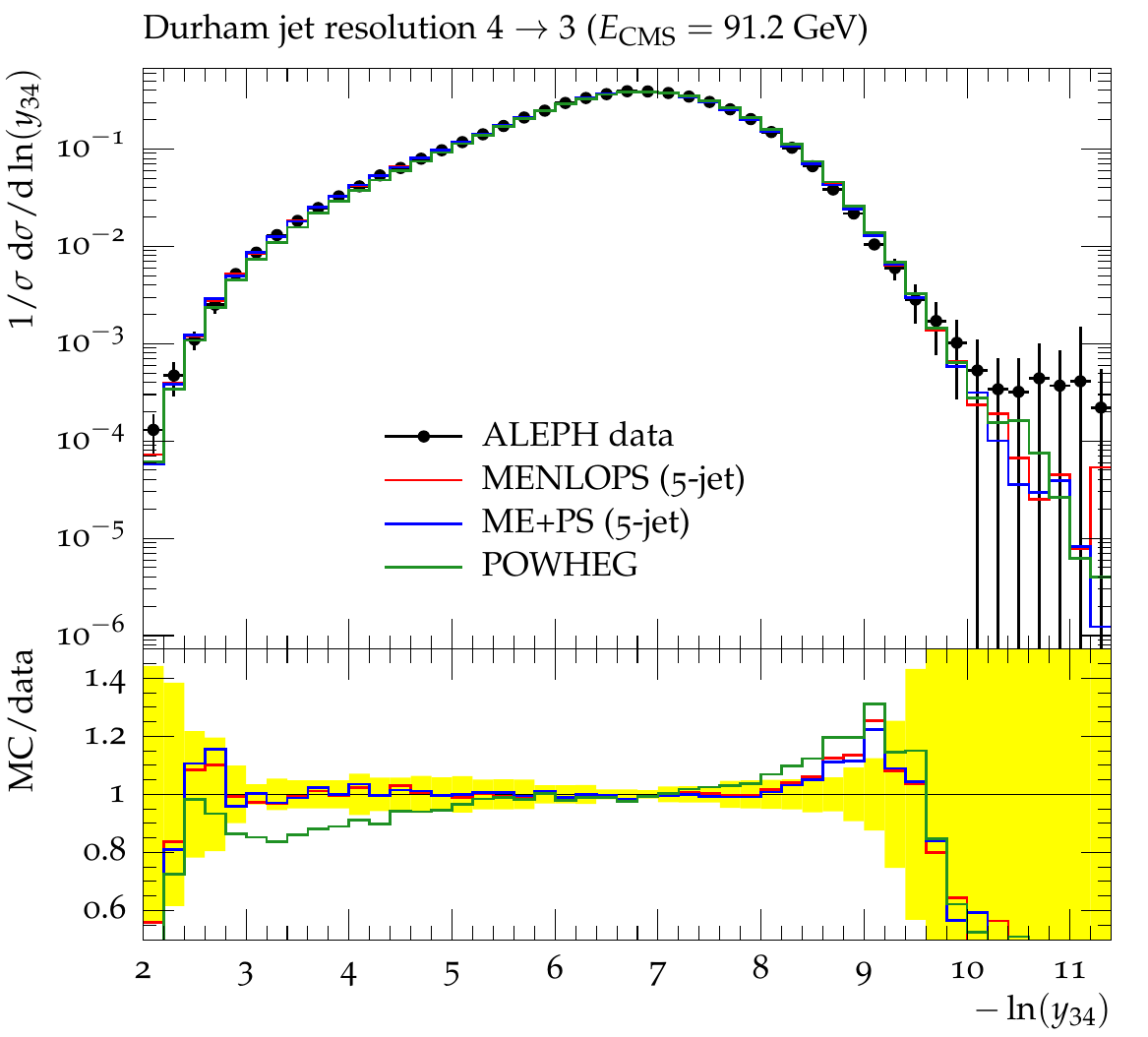}\hspace{20mm}
      \includegraphics{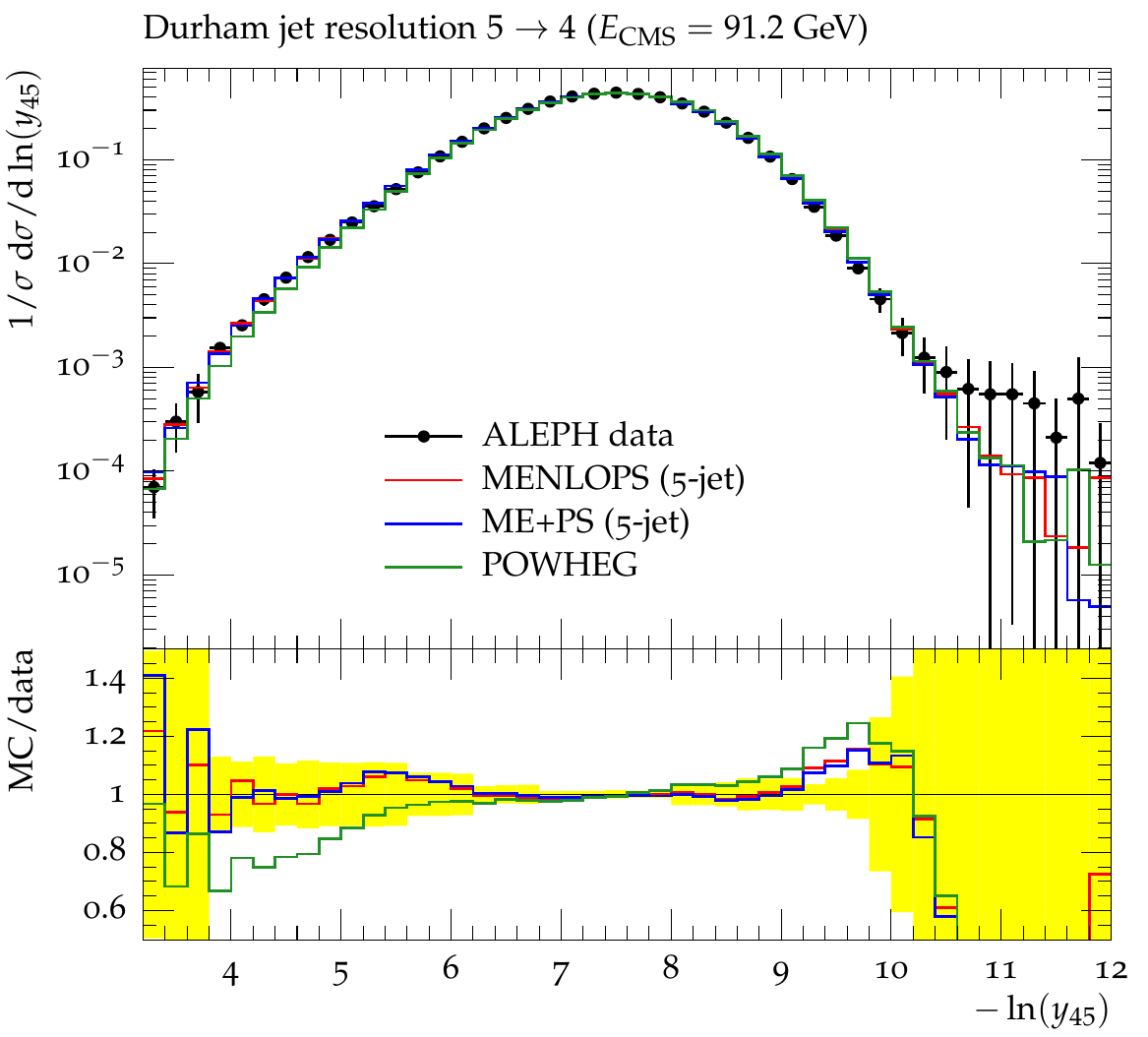}}
    \caption{Differential Durham jet rates as predicted by \Sherpa\
      using three different matching schemes, \POWHEG, \METS\ (up to 5
      jets) and \MENLOPS\ (up to 5 jets). The predictions are compared
      with LEP1 data measured by ALEPH in $e^+e^-\to$~hadrons
      \cite{Heister:2003aj}.}
    \label{fig:lep.menlops}
  \end{center}
  \vskip-5mm
\end{figure}

With the ME+PS facilities in \Sherpa\ at hand, it suggests itself to aim
at fusing the \POWHEG\ and ME+PS approaches. This effort goes under the
name \MENLOPS, and its key idea is to slice the \POWHEG\ phase space in
ME+PS style into two domains, the NLO core process domain and the
multi-jet domain. \MENLOPS\ has been developed very recently by two
groups as documented in Ref.~\cite{Hamilton:2010wh} and
Ref.~\cite{Hoche:2010kg}. The method exhibits what is cutting edge in
combining higher-order calculations with PSs. To understand the
slicing into domains, we schematically write down the expression for
an observable $\langle O\rangle$ in the \MENLOPS\ scheme:
\begin{eqnarray}\label{eq:menlops.obs}
  \langle O\rangle&=&\int d\Phi_B\,\bar B(\Phi_B)\left[\;
    \underbrace{\Delta^{(\mathrm{ME})}(t_0,\mu^2)\,O(\Phi_B)}_\mathrm{unresolved}\;+\,
    \sum_{ij,k}\frac{1}{16\pi^2}\int\limits^{\mu^2}_{t_0} dt
    \int\limits^{z_+}_{z_-} dz\int\limits^{2\pi}_0\frac{d\phi}{2\pi}
    \ \frac{R_{ij,k}(\Phi_R)}{B(\Phi_B)}\;O(\Phi_R)\right.
    \nonumber\\[2mm]&\times&\left.\bigg(
      \underbrace{\Delta^{(\mathrm{ME})}(t,\mu^2)\,
	\Theta(Q_\mathrm{cut}-Q_{ij,k})}_\mathrm{resolved,\ PS\ domain}\;+\;
      \underbrace{\Delta^{(\mathrm{PS})}(t,\mu^2)\,
	\Theta(Q_{ij,k}-Q_\mathrm{cut})}_\mathrm{resolved,\ ME\ domain}
      \bigg)\,\vphantom{\int\limits^{2\pi}_0}\right]
\end{eqnarray}
where $\Phi_B$, $\mu^2$ and $t_0$ denote the Born phase space,
factorization/high scale and infrared cut-off, respectively. The
one-particle emission phase space $\Phi_{R|B}$ is written explicitly,
the sum is over the relevant CS dipoles, and the \POWHEG\ and
\CSshower\ Sudakov form factors (no-branching probabilities),
$\Delta^{(\mathrm{ME})}$ and $\Delta^{(\mathrm{PS})}$, differ from
each other by using the $R/B$\/ and \CSshower\ kernels, respectively.
The domains can be read off Eq.~(\ref{eq:menlops.obs}) pretty
conveniently. The PS (or \POWHEG) domain is restricted to no resolved
and soft emissions ($Q<Q_\mathrm{cut}$) preserving the NLO accuracy
for inclusive observables. The hard (higher-order) emissions
($Q>Q_\mathrm{cut}$) are described by the ME (or ME+PS) domain
guaranteeing the LO+LL accuracy of each resolved jet emission. Note
that before fusing the contributions, the ME+PS part has to be
multiplied by the $K$-factor $\bar B(\Phi_B)/B(\Phi_B)$, as shown in
Eq.~(\ref{eq:menlops.obs}). In \Sherpa\ this $K$-factor is applied
locally, i.e.\ on an event-by-event basis.

\begin{figure}[t!]
  \begin{center}
    \resizebox{0.37\textwidth}{!}{
      \includegraphics{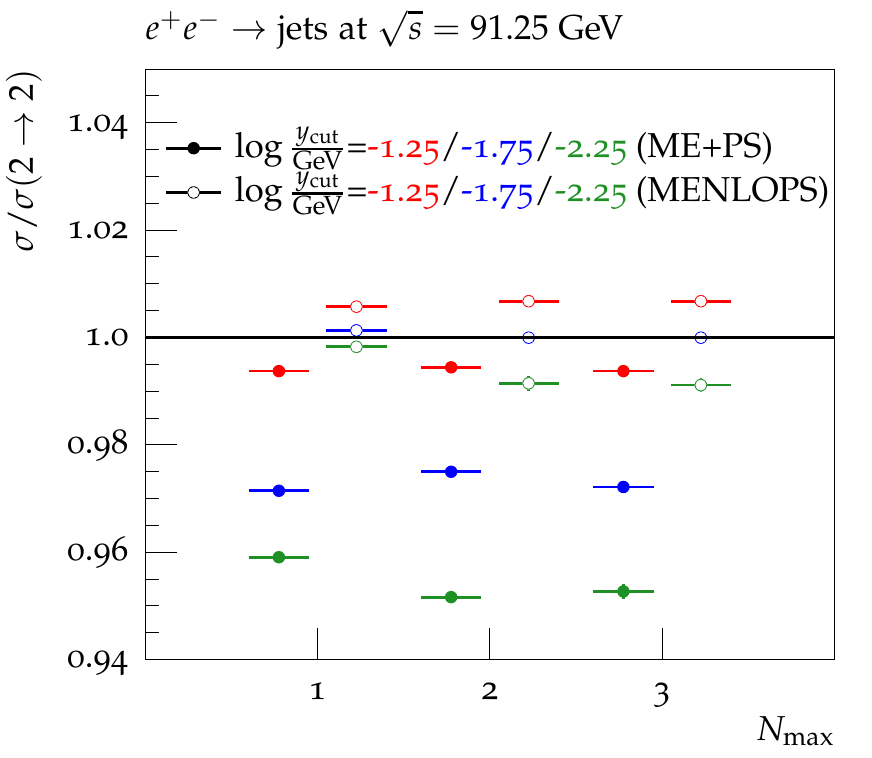}}
    \caption{Parameter dependence of the total inclusive cross section
      as predicted by \METS\ and \MENLOPS\ in \Sherpa\ for $e^+e^-\to$
      jets at LEP1. The cross sections are shown as a function of
      $N_\mathrm{max}$ for 3 different values of the merging cut.
      Note that the \METS\ and \MENLOPS\ results are normalized to the
      LO and NLO cross sections, respectively. $N_\mathrm{max}$
      denotes up to which multiplicity ($n+2$)-parton MEs are taken
      into account.}
    \label{fig:xsec}
  \end{center}
  \vskip-5mm
\end{figure}

\MENLOPS\ hence inherits the good features of NLO+PS and ME+PS, which
we demonstrate in Figs.~\ref{fig:lep.menlops}~and~\ref{fig:xsec}.%
\footnote{The NLO predictions shown in these plots were obtained by
  using virtual MEs provided by \textsc{BlackHat}~\cite{Berger:2008sj}.}
The differential jet rates for $e^+e^-\to$~ hadrons in
Fig.~\ref{fig:lep.menlops} prove that the shapes of \MENLOPS\ and \METS\
essentially are identical, and in very good agreement with the
data~\cite{Heister:2003aj} over the entire perturbative regime (which
is to the left in these plots). In contrast the \POWHEG\ predictions
fall short in describing the region of hard multiple jets. We display
in Fig.~\ref{fig:xsec} the parameter dependence of the \MENLOPS\ and
\METS\ total inclusive cross sections to show that NLO accuracy for
the core process leads to a NLO-like correction and stabilization of
the \MENLOPS\ cross sections. In the \POWHEG\ case
($Q_\mathrm{cut}\to\infty$) we were to find that the term in the
square bracket of Eq.~(\ref{eq:menlops.obs}) would integrate to one,
much like as in the pure parton shower case. The phase-space slicing
in ME+PS and in \MENLOPS\ necessarily generates a mismatch in the
non-logarithmic structure as given by the bracket term resulting in
deviations from the LO~(ME+PS) and NLO~(\MENLOPS) cross
sections.\footnote{The ``unitarity violations'' indicate the potential
  size of beyond NLO corrections; note that the pure \POWHEG\
  phase-space slicing effect is shown for $N_\mathrm{max}=1$.}
As shown in Fig.~\ref{fig:xsec}, the parameter dependence of the
\MENLOPS\ cross section is smaller maintaining the NLO accuracy almost
completely. This is where \MENLOPS\ improves over ME+PS.

\section{Conclusions and outlook\label{sec:conclude}}
Parton showers have been continuously improved and modernized over the
last years. The demand for improvement has come from measurements
reaching (for) higher precision at current hadron and future linear
colliders. The feasibility to aim at improvement came with the
fantastic advances in efficiently computing multi-leg tree-level and
one-loop amplitudes, including their integration over phase space. Two
directions have been established for systematically enhancing the
capabilities of parton showers:
\begin{enumerate}
\item Parton showers are improved by merging them with real-emission
  matrix elements for hard radiation (ME+PS). This is the new standard
  in the LHC era. \textsc{Alpgen}~\cite{Mangano:2002ea},
  \textsc{MadGraph/Event}~\cite{Alwall:2007st,Alwall:2011uj} and
  \Sherpa\ are widely used. The new ME+PS scheme in \Sherpa, \METS\
  (available since versions 1.2), greatly helped reduce the systematic
  uncertainties of older \Sherpa\ predictions. When compared to data,
  ME+PS predictions describe plenty of the measured shapes enabling
  the application of global $K$-factors that can be determined by
  higher-order calculations of the total inclusive cross section or
  the measurements themselves.
\item Parton showers are improved by matching them with NLO
  calculations (NLO+PS). \POWHEG~\cite{Alioli:2010xd,Alioli:2011nr,Barze:2012tt}
  and \MCNLO~\cite{Torrielli:2010aw,Frixione:2010ra,Frixione:2010wd}
  have a number of processes available.\footnote{Similar/Alternative
    approaches have been presented by other groups,
    e.g.~\cite{LatundeDada:2006gx,Giele:2007di,Hamilton:2008pd,Kardos:2011qa,Giele:2011cb,Platzer:2011bc,Alioli:2011as,Skrzypek:2011zw}.}
  For the latter, a\MCNLO~\cite{Hirschi:2011pa,Frederix:2011zi}, the
  new, automated \MCNLO\ framework developed by Frixione et al.\ in
  principle allows for tackling arbitrary processes provided the
  necessary amount of computer resources is available. \Sherpa's
  NLO+PS effort has been re-directed towards a \MCNLO-like strategy
  for many reasons; after gaining experience using a \POWHEG-like
  method~\cite{Hoche:2010pf,Hoche:2010kg}, it became clear that among
  other things a \MCNLO-like technique allows for much better control
  of the exponentiated terms, cf.~\cite{Hoeche:2011fd,Hoeche:2012ft}.
\end{enumerate}
Both directions are very active fields of research, and \MENLOPS\
actually emerged as a first successful attempt in fusing NLO
calculations with tree-level higher-order matrix elements. While
\MCNLO\ and \POWHEG\ give shower predictions of improved accuracy in the
basic process, \MENLOPS\ and ME+PS give improved multi-jet predictions.
\MENLOPS\ capabilities are enhanced over ME+PS regarding stability and
accuracy of the total inclusive cross section.

The frontier has been pushed as documented by many recent
publications~\cite{Alioli:2010xa,Melia:2011gk,Oleari:2011ey,D'Errico:2011sd,Frederix:2011qg,Garzelli:2011vp,Frederix:2011ig,Hoeche:2012ft,Campbell:2012am};
for example, NLO+PS techniques were applied to calculate $W+2$ and
$W+3$ jets. The former result was computed by
a\MCNLO~\cite{Frederix:2011ig}, while the latter result is a
documentation of the remarkable capabilities of the \MCNLO\
implementation that has become available in \Sherpa\
lately~\cite{Hoeche:2012ft}, provided efficient ``one-loop engines''
like \textsc{BlackHat} are interfaced as done in this $W+3$-jet
\Sherpa\ computation.

The above examples clearly demonstrate that NLO+PS for multi-jet final
states is no magic anymore, it is doable owing to the advances in NLO
calculations in the multi-jet realm. This actually brings ME+PS@NLO
within reach. First proposals have already appeared in the literature
\cite{Lavesson:2008ah}. The naive combination in form of NLO Exclusive Sums
discussed for $W+0,\ldots,4$ jets in \cite{Maestre:2012vp} has been
shown to work surprisingly well. To go towards ME+PS@NLO, it will be
necessary to replace each naive Exclusive-Sums jet veto at the
respective NLO accuracy by a jet veto at least accurate at
${\cal O}(\alpha^{m+1}_s)$ where $m$\/ is the highest LO jet
multiplicity (i.e.\ $m=4$ in the above example).

No matter which of these methods is finally used for phenomenological
studies, in all cases it is absolutely crucial to be able to provide
reliable estimates of the theoretical uncertainties of the
calculations. Comparisons between N(N)LO, NLO+PS, ME+PS, \MENLOPS\ are
mandatory to broaden our understanding regarding these
issues~\cite{Maestre:2012vp}.

\section*{Acknowledgments}
JW thanks the organizers for having created an excellent workshop
experience. SH's work is supported by the US Department of Energy
under contract DE-AC02-76SF00515, and in part by the US National
Science Foundation, Grant NSF-PHY-0705682, (The LHC Theory
Initiative). MS's work is supported by the Research Executive Agency
(REA) of the European Union under the Grant Agreement number
PITN-GA-2010-264564 (LHCPhenoNet). FS's is supported by the German
Research Foundation (DFG) via Grant DI 784/2-1.


\end{document}